# Scaling of Erosion Rate in Subsonic Jet Experiments and Apollo Lunar Module Landings

P.T. Metzger[1], John E. Lane[2], C.D. Immer[3], J.N. Gamsky[4], W. Hauslein[5],
Xiaoyi Li[6], R.C. Latta III[7], C.M. Donaue[8]

[1]Granular Mechanics and Regolith Operations Lab, NASA, Kennedy Space Center, FL 32899, Philip.T.Metzger@nasa.gov
[2]ASRC Aerospace, Kennedy Space Center, FL 32899, John.E.Lane@nasa.gov
[3]ASRC Aerospace, Kennedy Space Center, FL 32899, Christopher.D.Immer@nasa.gov
[4] Department of Physics and Astronomy, University of Kentucky, Lexington, KY 40506
[5] Department of Chemical Engineering, University of Mississippi, 134 Anderson Hall Post Office Box 1848, University, MS 38677-1848
[6] Code 548, NASA Wallops Flight Facility, VA 23395, Xiaoyi.Li-1@nasa.gov
[7] Department of Aerospace Engineering, Embry-Riddle Aeronautical University, Daytona Beach, Florida 32114
[8] Department of Physics, University of Colorado, Boulder, CO 80309-0390

## ABSTRACT

Small scale jet-induced erosion experiments are useful for identifying the scaling of erosion with respect to the various physical parameters (gravity, grain size, gas velocity, gas density, grain density, etc.), and because they provide a data set for benchmarking numerical flow codes. We have performed experiments varying the physical parameters listed above (e.g., gravity was varied in reduced gravity aircraft flights). In all these experiments, a subsonic jet of gas impinges vertically on a bed of sand or lunar soil simulant forming a localized scour hole beneath the jet. Videography captures the erosion and scour hole formation processes, and analysis of these videos post-test identifies the scaling of these processes. This has produced important new insights into the physics of erosion. Based on these insights, we have developed an erosion rate model that can be applied to generalized situations, such as the erosion of soil beneath a horizontal gas flow on a planetary surface. This is important to lunar exploration because the rate of erosion beneath the rocket exhaust plume of a landing spacecraft will determine the amount of sand-blasting damage that can be inflicted upon surrounding hardware. Although the rocket exhaust plume at the exit of the nozzle is supersonic, the boundary layer on the lunar surface where erosion occurs is subsonic. The model has been benchmarked through comparison with the Apollo landing videos, which show the blowing lunar soil, and computational fluid dynamics simulations of those landings.

## INTRODUCTION

A rocket exhaust with sufficient thrust to propel a lunar or martian lander also has the capacity to propel rocks, sand and dust, causing damage to any nearby hardware such as a lunar outpost. Simulations have shown that in lunar landings the velocity of the sprayed sand is on the order of 1 km/s (Lane, et al, 2008), which is 20 times higher velocity or 400 times higher particle kinetic energy than in a typical sandblasting operation. Analysis of human-sized landers on Mars shows that deep cratering will occur beneath the exhaust plume, ejecting soil and gravel into high trajectories that will rain down over a blast zone on the order of 1 km radius (Metzger, et al, 2009b). To develop mitigation strategies and technologies it is necessary to accurately predict these effects. Some of the physics is not well understood because there are not many situations in industry or daily life in which supersonic jets impinge upon granular media, especially considering the extreme environments: different ambient atmospheres and gravity than on Earth, and unusual soil.

**SMALL-SCALE EXPERIMENTS**

**Relevance to Lunar Case.** It is very difficult to perform an experiment that captures the correct physics for rocket exhaust blowing lunar soil. Because the rocket exhaust expands into the lunar vacuum, it is rather rarefied (as shown by simulations) in the region where soil erosion occurs, with the Knudsen number being on the order of unity relative to a sand-sized particle. On Earth, this can be re-created in a vacuum chamber, but as soon as the rocket engine is ignited, the plume will fill the chamber and it will not be a vacuum. In an atmosphere the rocket exhaust is collimated into a jet and does not produce a realistic pressure or shear stress profile on the soil. Vacuum chambers do exist with adequate pumping capacity to keep up with a full-scale engine, but such chambers too expensive for the lower-cost experiments that we need earlier in the program, and the pumps would be ruined by exposure to the blowing lunar dust. Large vacuum chambers can hold a low and reasonably constant background pressure without pumping for a short duration test if the simulated rocket plume is scaled very small, but unfortunately there are too many aspects of the physics involving the inherent length scale of the sand grains (permeability, Knudsen number, van der Waals cohesive forces, etc.) to permit size scaling without compromising some aspect of the physics. Granular physicists have come to the consensus over the past couple of decades that the only consistent scale is full-scale; see for example Orpe and Khakhar (2001). Unless we already fully understand the physics, we cannot be sure what aspects we are compromising and how that will affect the outcome of the experiments relative to what occurs in the real, full-scale, lunar case. Furthermore, it is difficult to perform reduced gravity experiments except for small scale in aircraft flying parabolic trajectories, and all the more difficult if we wanted to install a sufficiently large vacuum chamber in the aircraft. Therefore, the only feasible way to investigate the physics of rocket exhaust blowing soil is to rely upon physics-based modeling. Experiments can be performed that are small-scale in vacuum or in low gravity, and larger scale at ambient atmosphere to calibrate and benchmark the model, data from planetary missions can also serve that purpose as they become available. As we gain confidence that the physics are modeled correctly, then we can rely on the model to make predictions for the large-scale, fully lunar case. Therefore, experiments can contribute to the physics of the model even if they are very unlike the lunar case, as long as they provide insight into the physics and/or provide a case that can be modeled for benchmarking against the experimental results.

Toward these ends, we have conducted experiments using subsonic and supersonic jets in a terrestrial atmosphere impinging upon beds of various sands or soils, including simulated lunar and Martian soil. Some of the experiments have been in reduced gravity or hyper-gravity. We have also developed an experimental apparatus for use in a large vacuum chamber to vary the Knudsen number, and we plan to complete these tests within another year. The subsonic experiments have been most useful for the lunar case to identify the scaling of erosion rate. Even though the rocket exhaust plume from a lunar lander is hyper-sonic, the flow in the boundary layer where erosion takes place is subsonic and so the scaling with gravity, Knudsen number, sand grain size, and other parameters in these experiments has provided significant insight.

**Subsonic Cratering.** The cratering apparatus has been described by Metzger, et al (2009a and 2009c). In brief, it consists of a box filled with sand, having a transparent front wall with a sharp, outwardly-beveled upper edge. The jet is produced by a straight, circular, vertical pipe that is centered over the sharp edge of the bevel. The circular jet exiting the pipe is thus split in half by that edge, with the semi-circular half of the jet that goes inside the box causing a crater to form in the sand adjacent to the window, so that the crater can be seen through the window as it forms. Video analysis software measures the crater width, depth, volume, and other parameters in every frame of the video of the experiment. The crater formation has been analyzed two different ways. First, the depth of the crater has been shown to have the form $d = a \operatorname{Ln}(bt+1)$ over a significant period of time, where *d* is depth, *t* is time, and *a* and *b* are fitting parameters. These can be interpreted as the length scale and inverse time scale, respectively. Since volume scales as the depth cubed, at least in the initial crater formation, the erosion rate scales as $a^3 b$. Second, the volumetric erosion rate can be measured directly. This rate is not constant because of the initial transients and because the crater growth slows as the widening crater begins to re-capture and re-circulate an increasing share of the ejected sand. Thus, we measure the peak erosion rate in each experiment, which follows after the transients are complete but before recirculation of sand begins to dominate. Either way, the experiment parameters can be varied (gas velocity, gas molecular weight, sand grain density, sand grain size, pipe diameter, and pipe exit plane height) so that the scaling of the erosion rate can be determined for each parameter.

**Reduced Gravity.** To obtain erosion rate scaling with gravity, we have constructed a totally enclosed version of the subsonic cratering apparatus. The apparatus uses a fan to force air into the pipe over the beveled window. The experiment was flown for seven flight days on reduced gravity aircraft, producing cratering data for 1/6 gee (lunar case), 3/8 gee (martian case), 1 gee (terrestrial case), and 2 gee (hyper-gravity). On each flight day a different soil was used, including quartz sand of several different grain sizes, four different lunar soil simulants, and a Mars soil simulant. On each successive reduced gravity parabola, the fan setting was adjusted to provide a different gas velocity. Gas density could not be controlled independently and was always determined by the aircraft cabin pressure. The experiment was small enough for a single person to lift, so between parabolas it could be tipped upside down and shaken side-to-side to replace the sand behind the viewing window and to re-level it.

**Supersonic Cratering.** Tests were also performed using small, 100 lbf solid rocket motors firing into a very large sand bed outdoors in order to observe the differences when the jet velocity is supersonic. We observed transient shock effects from motor ignition, and deep, violent cratering in the sand. The cratering was dominated by the bearing capacity failure mechanism (Alexander, et al, 1966) due to the extremely high stagnation pressure of the impinging jet. This is very unlike the lunar case seen in the Apollo and Surveyor landings, and is more relevant to the martian case (Metzger 2009b).

**Experimental Results.** Figure 1 shows several cross-sectional profiles of a typical crater at intervals during its growth beneath the subsonic jet in 1 gee. At first it is approximately paraboloidal, but after reaching a characteristic size the gas traction holding up the sides of the crater becomes insufficient and the top edge collapses to form an outer slope at the angle of repose specific for this material (Metzger, et al, 2009a). This outer crater only forms for higher velocity jets; for slower jets the crater is broader and shallower, with a slope below the angle of repose. The paraboloidal "inner" crater then remains approximately constant while the outer crater grows as an inverted cone of constant slope. By fitting a polynomial onto the inner crater profile and a straight line at the angle of repose onto the outer crater profile, it is possible to obtain the relationship between depth and width (at the surface) for this family of curves. The integral of rotation for this relationship produces an analytical function $V(d)$ where $V$ is volume and $d$ is depth. The video analysis software that we created also computes the integral of rotation for the crater profile but does so numerically without the analytical functions, and when we examined the empirical depth vs. volume data (e.g. figure 2 for a 10 m/s jet) we discovered that an exponential $V(d) = C \cdot e^{Fd}$, where $C$ and $F$ are fitting parameters, is adequate for the purposes of extracting scaling relationships (apart from an initial transient of short duration). Apparently the analytical $V(d)$ happens to approximate the Taylor series of an exponential function. (In most cases, the slope of the log-log plot for volume versus time changes abruptly just as the outer crater begins to form. This is because the depth to volume relationship is different for paraboloids and cones. In those cases, the analysis could focus on just the inner or just the outer crater or both) Furthermore, we find empirically that the volume grows as a power function $V = A \cdot t^B$. Equating these two expressions for $V$ and rearranging obtains $d = a \operatorname{Ln}(bt)$, where $a = B/F$ is the length scale and $b = (A/C)^{1/B}$ is the inverse time scale. Because the $V(d)$ relationship is not valid for the initial transient, we prefer to write $d = a \operatorname{Ln}(bt+1)$ so that $d=0$ when $t=0$, and we find that this fits better during the initial transient.

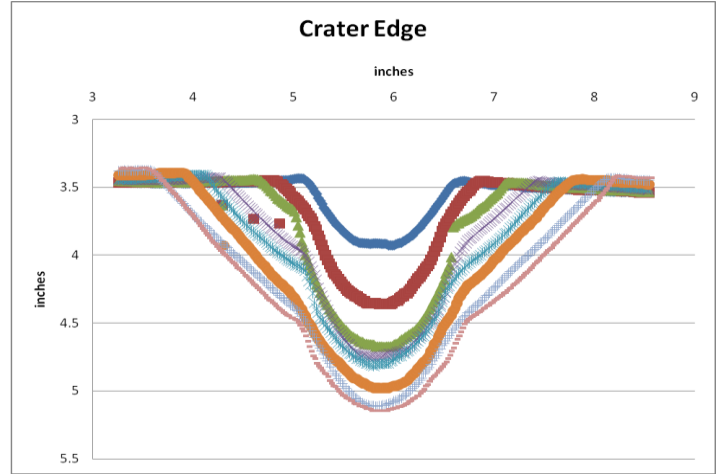

**Figure 1.** Crater profiles measured at the viewing window.

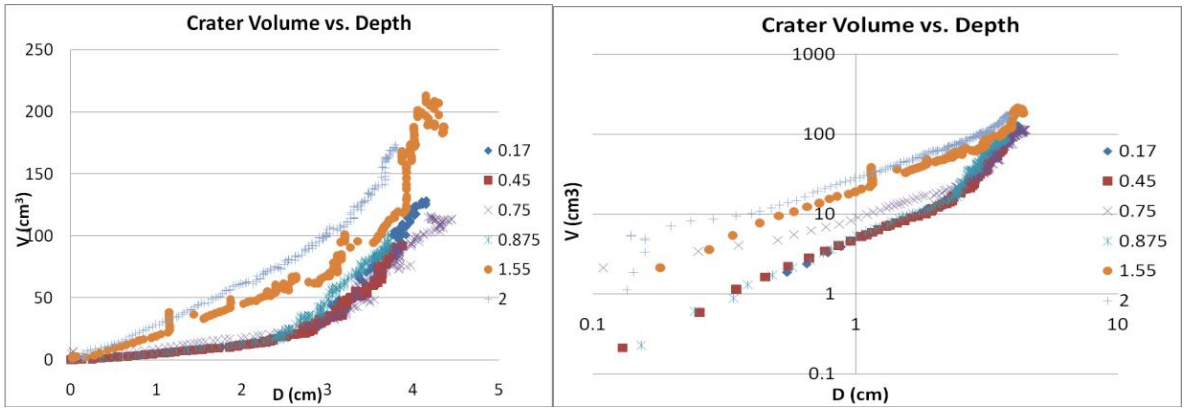

**Figure 2. Volume versus Depth. Left: linear plot. Right: log-log plot.**

Our earlier work (Metzger, et al, 2009a) indicated that the erosion rate scaling goes as:

$$a \propto H \qquad b \propto \frac{\rho v^2 D}{H^3} \qquad (1)$$

where $H$ is the jet height, $D$ is the jet diameter, $\rho$ is the gas density, and $v$ is the jet velocity. However, we also note that the dependence of $b$ upon $D$ may have been nearly as consistent with $D^2$ instead of $D$, and is within the uncertainty of the experimental data, and that on theoretical grounds that would be preferred, as follows. The volumetric erosion rate is

$$a^3 b \propto \rho v^2 D^\beta \qquad (2)$$

where remarkably there is no *H* dependence. This is reasonable because within a range of *H* all the jet energy is still delivered to the soil. The crater size scale *a* increases linearly with *H* because the jet widens linearly with distance from its exit plane. The growth of depth should slow and thus *b* (the inverse time scale) decreases with *H* so that the duration to reach a given depth must increase. But as long as *H* is not too large, all the jet energy is still delivered to the soil and thus overall volumetric erosion rate should not be affected. Thus $a^3 b$ should be independent of *H* and so *b* should vary as $H^{-3}$. This argument is centered on the idea that erosion rate scales with delivery of energy to the soil. If that is the case, then *b* should scale as $D^2$ since that is proportional to the area of the jet.

More recently, Metzger, *et al*, (2009c) reported erosion rate scales with particle diameter *d* as $a \propto d^{1/2}$ and $b \propto d^{-2}$. However, we now believe on theoretical grounds that $a \propto d^{1/3}$ and $b \propto d^{-2}$ is probably the correct scaling so that $a^3 b \propto 1/d$. Figs. 3 and 4 show the scaling of *b* and *a* respectively with particle diameter. The fit for $a \propto d^{1/2}$ is marginally better than $a \propto d^{1/3}$, but both are within experimental uncertainty. We note that in Eq. (3) the scaling with jet parameters is proportional to the densimetric Froude number squared. If we choose the 1/3 exponent for *a*, then the overall dependence is

$$a^3 b \propto \frac{\rho v^2 D^2}{(\rho_g - \rho)^\gamma g^\varepsilon d} \approx \frac{\rho v^2 D^2}{\rho_g^\gamma g^\varepsilon d} \tag{3}$$

where $\rho_g$ is the density of the grain material, *g* is gravitational acceleration, and exponents γ and ε are expected to both be unity for consistency with the squared densimetric Froude number. Indeed, our more recent experimental data shown in figure 5 have shown that both exponents are consistent with unity. (Exponents of 1.3 provide a marginally better data collapse in figure 5, but on theoretical grounds unity appears to be the better conclusion for consistency with the scaling in *d* in Eq. 3.) The left and right sides of Eq. 3 still do not form a dimensionless group. It is tempting to make *v* cubed in the numerator to represent energy transport and achieve consistency in the units. However, the dependency on *v* in Metzger (2009a) was clearly squared, not cubed, and so another velocity is needed to complete the group. An example of a velocity with physical relevance is v(*d*), the asymptotic velocity of an airborne particle in the gas flow, proposed by Roberts (1967) to play the mediating role between shear stress and erosion rate. There may be other candidates. In any case, we conclude that the erosion rate is proportional to the *squared* densimetric Froude number. In the following section, this result will be related to the erosion rate of soil in the Apollo landings as measured through the optical density of the blowing clouds visible in the landing videos.

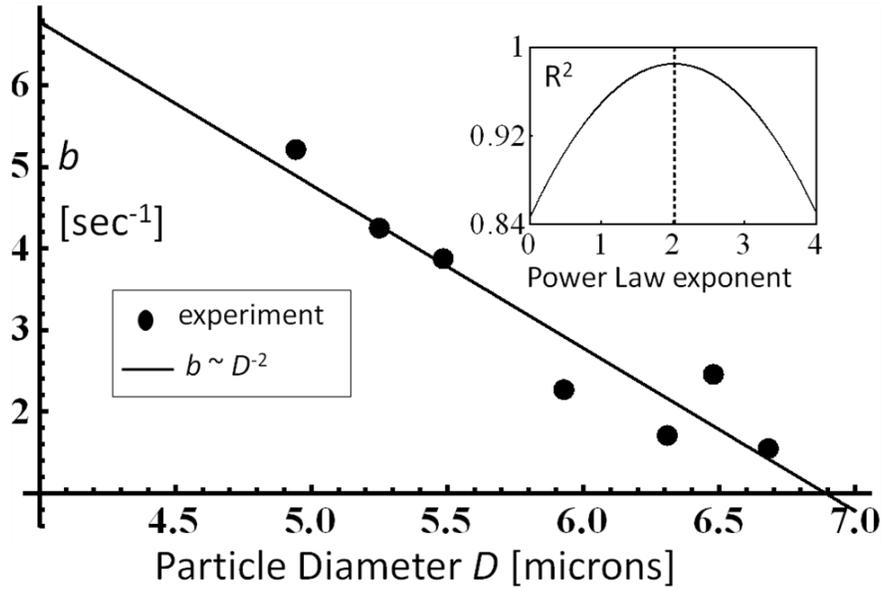

**Figure 3. Dependence of length scale *b* with particle diameter *D*.**

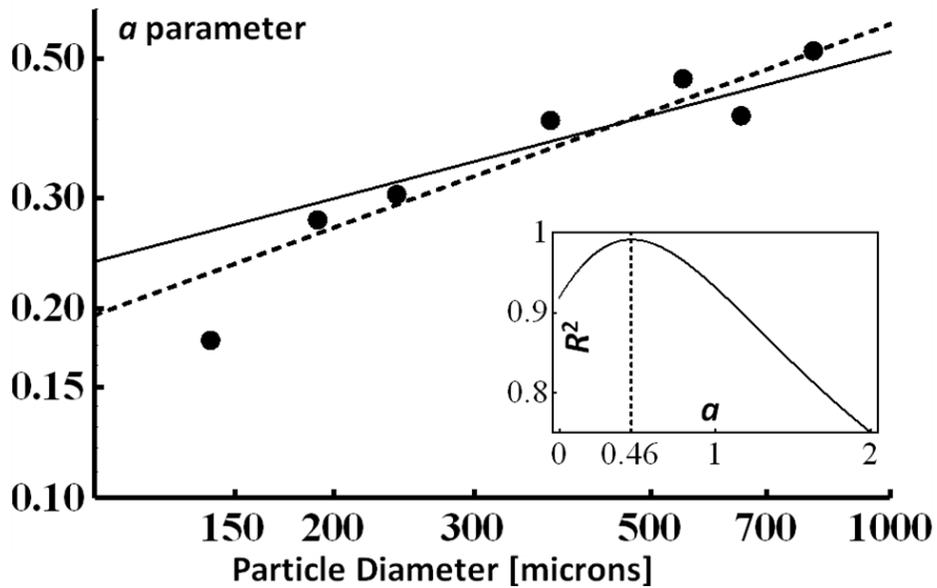

**Figure 4. Dependence of length scale *a* with particle diameter *D*. Solid line is $D^{1/3}$. Dashed line is $D^{1/2}$. Optimum is $D^{0.46}$.**

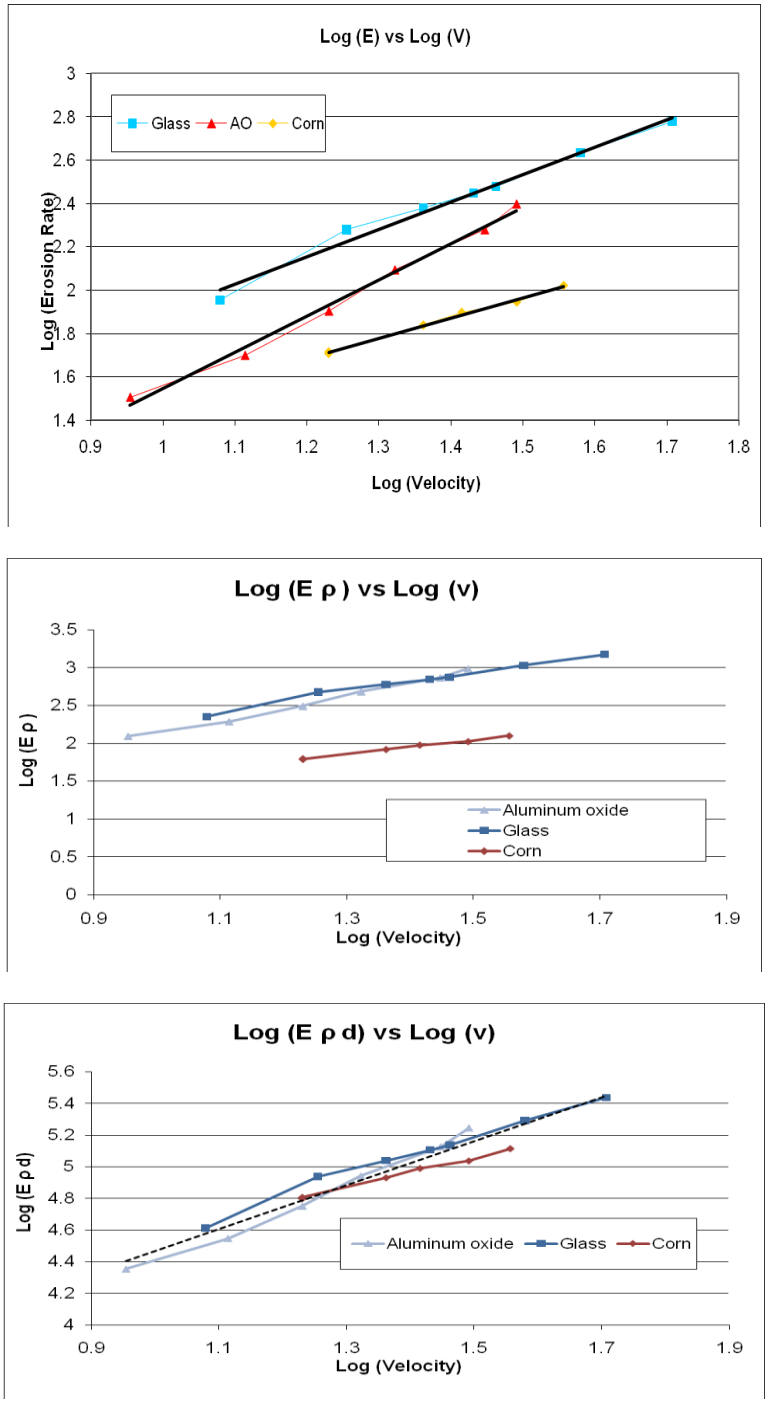

**Figure 5. Dependence of volumetric erosion rate upon grain density and particle diameter. Top pane: No data collapse with grain density and diameter are ignored. Middle pane: The two materials with the same grain sizes but different densities collapse when density is accounted for. Lower pane: all three materials collapse when density and diameter are accounted for.**

**EROSION RATE IN APOLLO LANDINGS**

The erosion rate of soil in extreme environments may also be measured directly from prior landings on the Moon or Mars. No instrument has been flown to directly measure these effects. The extant data sets are (1) photographs of the soil beneath the Surveyor landers on the Moon, Viking landers on Mars, Apollo Lunar Modules (LM) on the Moon, and the Phoenix lander on Mars, (2) hardware from the Surveyor III spacecraft brought back by the Apollo 12 astronauts and demonstrating the damage it sustained from its own landing and from the Apollo LM landing, (3) and the Apollo LM landing videos.

The photographs taken to-date are not useful for calibrating erosion rate in rocket landings. Mason and Nordmeyer (1969) and Mason (1970) have assessed photographs of putative craters beneath a Surveyor lander nozzle in comparison with terrestrial experiments to derive an erosion equation. The difficulties with this approach are that we cannot be sure which (if either) of the craters was caused by the Surveyor thrust, and that even if they were then the erosion due to the ignition overpressure (shock effects of ignition) of those thrusters are largely unrelated to the scaling of erosion of a descending spacecraft without those transient shock effects. Moore, et al (1987) studied the surface beneath the Viking thrusters and judged that numerous indications of plume interactions were visible. Because of the complexity of the types of soil (clods, crusts, etc.) and the uncertainty of the plume conditions when the lander is low to the ground and recirculation (in the Martian atmosphere, with multi-engine propulsion) it is difficult to extract an erosion rate from these images. Metzger, et al (2009b) studied supersonic exhausts impinging on soil and concluded that crater growth is governed by jet length and not by the rate of viscous erosion integrated over time, so the Viking results are not useful to calibrate viscous erosion rate. The soil beneath the Apollo landings generally shows no crater, presumably because the soil was eroded over such a wide diameter to such a shallow depth that the change in surface topography cannot be identified. One possible exception is a crater identified in the Apollo 14 mission (Katzan and Edwards, 1991), but we see no way to calibrate an erosion rate from its volume since the LM was not stationary over it and we do not know the dwell time. Mehta, et al (2009) have studied the removal of soil by the pulse modulated Phoenix thrusters and have concluded that it was driven by cyclic shock effects enhancing gas diffusion into and fluidization of the soil. That mechanism is very different than the ordinary viscous erosion being studied here.

We have recently concluded that the Surveyor hardware is not useful for calibrating a soil erosion rate because that spacecraft was probably located beneath the main sheet of spraying soil from the Apollo LM and thus was exposed to only the fringes of the spray (Immer, et al, 2009). That is because the Surveyor was down inside a crater at a lower elevation than the Apollo 12 LM, and unless the local terrain of the LM just happened to tilt down to point exactly toward the Surveyor, then the thin sheet of spraying soil would miss the Surveyor. The sheet of soil has an angular thickness of about 3 degrees (Immer, et al, 2008), so the tilt of the local terrain would have to be accurate to within ±1.5 degrees. Our optical density calculations (below) compared to the damage of the Surveyor, implies that this was not the case as far less damage occurred to the Surveyor than the direct sheet would have caused.

**Optical Density of Blowing Lunar Soil.** The only method we have, then, to calibrate the erosion rate from actual spacecraft landings is to measure the optical density of the blowing dust clouds visible in the Apollo landing videos. As shown in figure 6, the optical density increases as the LM is closer to the ground. We have measured the optical density using the method described by Immer, et al (2008). In that reference, the calculation of mass density in the cloud was unrealistic because the particle size distribution did not account for the large number of submicron fines, which we show below to be important. At a "typical" point in the plume (used in the following calculations), T=32.35% of the incident light is transmitted through the cloud. This is related $N$, to the number density of particles in the cloud, by

$$T = \exp\left\{-N s \int_0^{600} \mathrm{d}d \int_{0.38}^{0.75} \mathrm{d}\lambda \left(\frac{\pi d^2}{4}\right) C_E(d,\lambda) S(\pi) P(d)\right\} \qquad (4)$$

where $s$ is the path length of light through the cloud, $d$ is particle diameter, $\lambda$ is wavelength, $C_E$ is the coefficient of extinction of light through a spherical particle, $S$ is the solar spectrum (normalized to unity over the interval of this integral, and taken to be flat in this approximation), and $P$ is the particle size distribution of particles in the cloud. Particle sizes are integrated from 0 to 600 microns (an estimate of the approximate largest particle to erode and/or to affect the calculations significantly). Wavelengths are integrated from 0.38 to 0.75 microns, the visible portion of the spectrum.

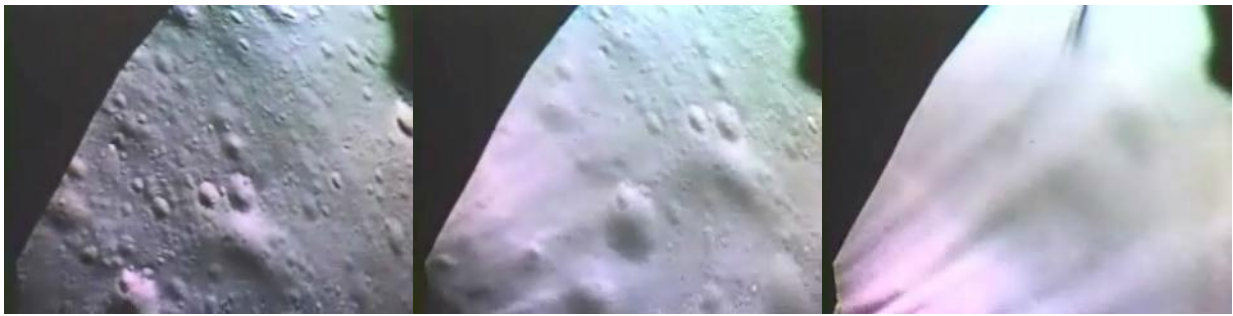

**Figure 6. Images from the Apollo 16 descent imaging camera showing increasing optical density of the dust cloud as the LM nears the surface.**

The extinction coefficient of lunar soil particles, figure 7 (right), is calculated as

$$C_E = \begin{cases} 4\alpha \,\text{Im}\left[\dfrac{(n-ik(\lambda))^2-1}{((n-ik(\lambda))+2)^2}\right] + \dfrac{8}{3}\alpha^4 \,\text{Re}\left[\dfrac{(n-ik(\lambda))^2-1}{((n-ik(\lambda))+2)^2}\right] &,\ \alpha\ \text{small} \\[2ex] 2 &,\ \alpha\ \text{big} \end{cases}$$

where the imaginary part of the index of refraction, $k$, is shown in figure 7 (left) (Lecy, 2009), and the real part, $n$, is taken to be 1.75, and $\alpha = \pi d/\lambda$ is a dimensionless particle size. Some assumptions are needed for the transition to the classical limit at big $\alpha$. We assumed the transition occurs when the top half of the equation equals 2 and applied an averaging function to smear across the transition. Varying these assumptions through reasonable bounds shows that the final prediction of eroded soil is affected by as much as a factor of 2, so some improvement is needed. This is, however, a dramatic improvement over estimates of erosion made from the Surveyor damage or from other existing methods.

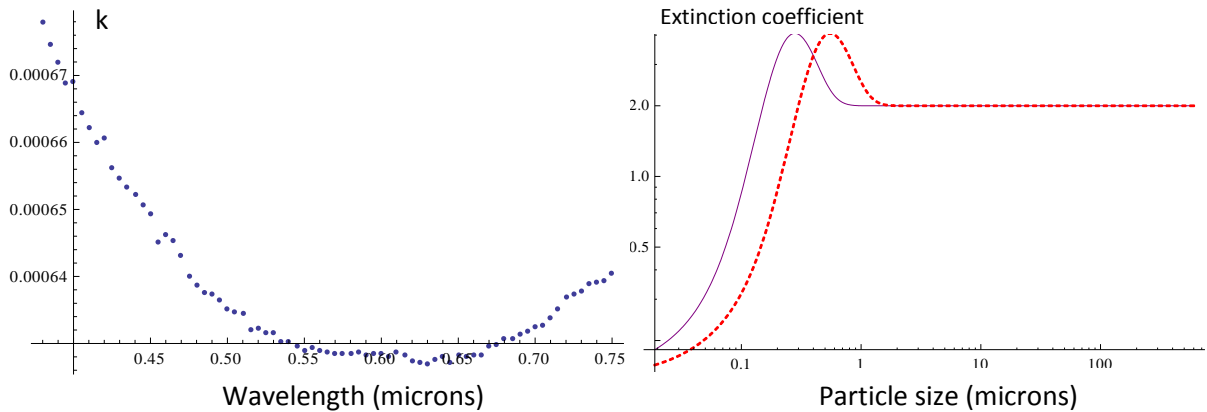

**Figure 7. (Left) k-spectrum of lunar soil. (Right) extinction coefficient of spheres of lunar soil materials: (purple) wavelength 0.38 μm; (red) wavelength 0.75 μm.**

**Lunar Soil Particle Size Distribution.** As shown above, the particle size distribution is needed to calculate the erosion rate from optical density. In Immer, et al (2008), we did not take into account the very large number of particles smaller than a micron. The following analysis corrects this. Figure 8 (left) shows the number density of the lunar simulant JSC-1A for particles larger than a micron as measured on a Fine Particle Analyzer, an optical instrument that has a lower measurement limit of about 1 micron. Figure 8 (right) shows this number density converted to a mass density (assuming the particles are spheres) showing how it may be approximated well by an analytical function consisting of two power laws with a transitional function between them. Figure 9 shows this function concatenated with each of three different particle size distributions for fines smaller than a micron as measured by Park, et al (2008). These three distributions are for two different lunar samples and for JSC-1AvF, the version of lunar simulant JSC-1A that includes only the very finest particles. In each case, the concatenation is weighted so that the fines below 10 microns constitute a prescribed fraction of the mass of the soil as predicted by the combined distribution. We have used fractions between 3% and 16% following the various lunar soils plotted in figure 9.1 by Carrier, et al (1991). The fraction of surface area (and the relative contribution the the optical density) attributable to these fines can be estimated by integrating any of these versions of the concatenated particle size distribution across all particle sizes with the weight factor ($\pi d^2/4$), the particle cross-sectional area assuming they are spheres. An example is shown in figure 10. Using an Apollo fines sample and assuming 8% of the soil by mass is below 10 microns, we find that approximately 2/3 of the surface area in the cloud is due to the particles below 10 microns, the range that was not measured accurately by our earlier fine particle analysis. This shows that previous estimates of erosion that had neglected the submicron particles had significantly over-estimated the mass needed to produce the measured optical densities. The following calculations are made using Apollo sample 10084 assuming the mid-range estimate of 8% by mass smaller than a micron.

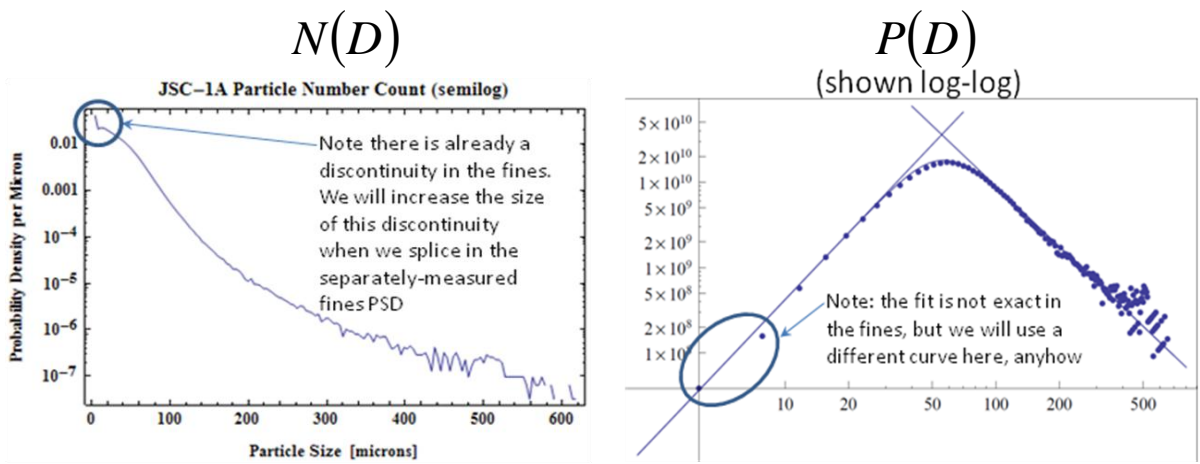

**Figure 8. (Left) Number density for JSC-1A. (Right) Particle size distribution for JSC-1A.**

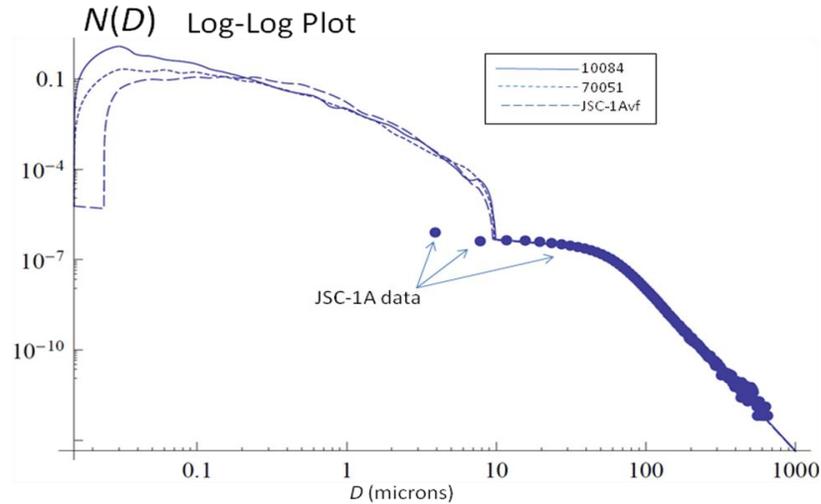

**Figure 9. Concatenated particle size distribution (log-log).**

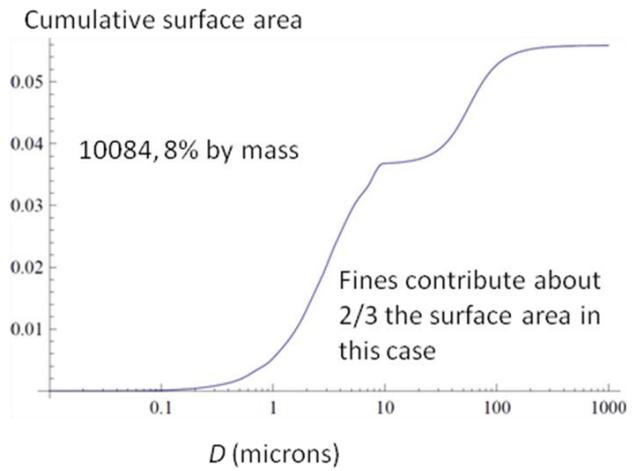

**Figure 10. Model of cumulative surface area of lunar soil (log-linear).**

Finally, we must note that the particle size distribution in the cloud is not identical to that of the soil lying on the ground, because particles of different sizes are accelerated to different velocities by the plume and thus are "stretched out" through space to different densities. The "blowing" distribution is calculated by dividing the "static" distribution by the particle velocities (as a function of their size) and multiplying by a renormalization constant (such that the total distribution still integrates to unity with the correct engineering units). The particle velocities as a function of $d$ were taken from numerical simulations by Lane, et al (2008). The static and blowing distributions are compared in figure 11. We used the blowing distribution in our calculations.

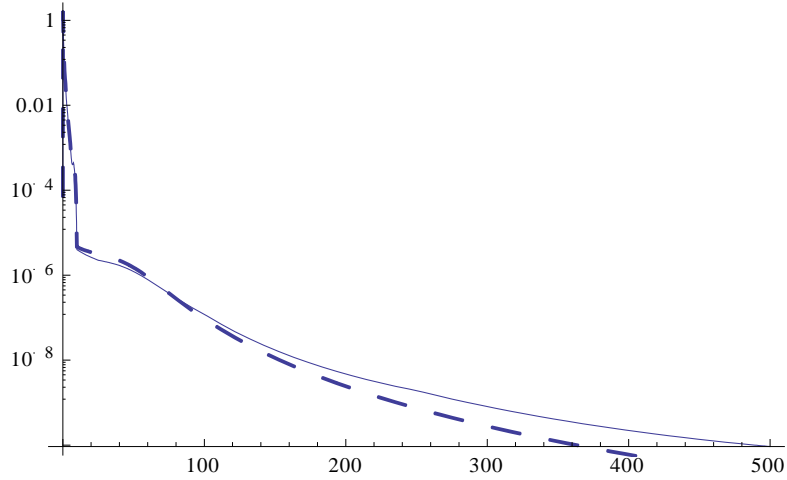

**Figure 11. Concatenated Particle Size Distribution (Semi-log). Solid: blowing soil. Dashed: static soil.**

## SEMI-EMPIRICAL EROSION EQUATION

**Form of the Erosion Equation.** Roberts (1963) has stated that the shear stress developed by a wall jet over a very rough surface is approximately the dynamic pressure of the jet, and in the case of a wall jet emanating from the impingement point of a lunar lander plume it scales with the thrust of the lunar lander $\rho v^2$. Our experiments show that this is also proportional to the erosion rate $\dot{m} \propto \rho v^2$ and hence erosion rate is proportional to shear stress. We can consider the erosion equation to be of the form $\dot{m} = \tau\, f(\text{soil, gravity})$, where $\tau$ is the local shear stress of the gas over the soil at a particular location and $f$ is a function that captures the dependence on soil particle size distribution, gravity and other parameters. If we assume the soil to be roughly the same over the entire Moon, then we can replace $f$ by a constant value determined by measuring $\dot{m}$ and $\tau$ at any one location and time. We cannot actually see the quantity of soil eroded at one point of the ground, but rather the soil that eroded at all points upstream of the observation. Hence we integrate radially,

$$\text{Flux} = \frac{1}{R^2 \tan\theta} \int_0^R \dot{m}\, r\, dr = \frac{f}{R^2 \tan\theta} \int_0^R \tau(r)\, r\, dr \qquad (5)$$

where $R$ is the distance at which the optical density is observed and $\theta$ is the angular thickness (vertically from the local terrain) of the blowing sheet of soil (used to calculate path length $s$ in

Eq. 4) as measured by Immer, et al (2008a) using shadows of the LM in the Apollo landing videos. The "Flux" is the mass per unit time that is crossing a cylindrical surface of radius $R$ and height $R \tan \theta$ centered on the plume impingement point. It is found by

$$\text{Flux} = N \int_0^{600} P(d) \, v(d) \, \rho_g \left( \frac{\pi d^3}{6} \right) \mathrm{d}d \qquad (6)$$

where $N$ was calculated by rearranging Eq. 4. Thus, combining (5) and (6),

$$f = \frac{(\pi/6) \rho_g N R^2 \tan\theta \int_0^{600} P(d) \, v(d) \, d^3 \, \mathrm{d}d}{\int_0^R \tau(r) \, r \, \mathrm{d}r} \qquad (7)$$

**Numerical Results.** We have obtained $\tau(r)$ numerically by performing several steady state Reynolds Averaged Navier Stokes (RANS) Computational Fluid Dynamics (CFD) simulations for the LM descent engine at altitudes of 20, 10 and 5 meters above the lunar surface. The CFD uses an unrealistically smooth surface to represent the lunar surface and so the shear stress is probably much smaller and more uniform than in a real landing. Roberts (1963) stated that the shear stress in a jet over a very rough surface may be approximated as the dynamic pressure of the jet. In our simulation, we find that the shear stress is many orders of magnitude weaker than the dynamic pressure and that they are not simply proportional. Their ratio varies over two orders of magnitude, including in the area where erosion should be most significant. Therefore, we recognize that much more work is needed to average shear stress across many simulations with differing, realistic terrain shapes, and to average over the optical density across larger sweeps of the visible dust cloud. This should obtain an averaged relationship between shear stress and optical density suitable for realistic lunar terrain. In the meantime, the following numbers based on an idealized CFD boundary are therefore to be regarded a *model* of lunar erosion useful in conjunction with similar CFD simulations. This model is an improvement over prior lunar erosion models but is not to be considered a physical erosion law.

Our calculations show that for landers at 20m and 10m height, the integral over $\tau$ (denominator of Eq. 7) evaluates to 17.4 mN and 35.9 mN, respectively. We linearly interpolate this to 25.0 mN at 14.14 m altitude, which is where one of our optical density measurements took place. In that measurement, $R= 66.26$ m, $\theta=2.3°$, and $N= 9.95\times10^{11}$ grains/m$^3$. Thus, we find $f=22,000$ kg/s/N=1/(4.55 cm/s). For similar terrestrial experiments using simulated lunar soil, this may be extrapolated to Earth's gravity as:

$$\dot{m}_{\text{Earth}} = \cdot \frac{g_{\text{Moon}}}{g_{\text{Earth}}} \times \frac{\tau_{CFD}}{4.55 \text{ cm/s}} = \frac{\tau_{CFD}}{27.5 \text{ cm/s}} \qquad (8)$$

We note that the velocities in the denominators of Eq. 8 play the role of the particle's asymptotic airborne velocity (*au*) as proposed by Roberts (1963), but they are much slower corresponding to the much lower values of $\tau$ in our CFD's, and they may not have the same physical interpretation. Roberts assumed that a particle accelerated to its asymptotic velocity over a very short distance, thus removing excess shear stress from the gas in the same locale from which it eroded and providing a "feedback" to control erosion rate. We do not see the particles reaching full velocity over such a short distance in our simulations, and so we think Roberts' interpretation of this velocity is probably not correct. We also note that a velocity was missing in the scaling relationship Eq. 3, and this is probably related to the velocity of Eq. 8. The question is: what is the correct physical interpretation of this velocity?

**CONCLUSIONS**

Simple erosion experiments using a subsonic jet have been useful to identify scaling relationships. It appears that the volumetric (or mass) erosion rate is proportional to the square of the densimetric Froude number and thus is proportional to shear stress and inversely proportional to gravity. Because soils are a complex mixture of many particle sizes we are not ready to scale erosion rate as a function of particle sizes for real soils. Our analysis has to-date also neglected the cohesion of the soil, which in the lunar environment with low gravity appears to be significant. The absolute erosion rate based upon the shear stress in a CFD can be calibrated by comparison with the optical density in the Apollo landing videos. This has identified a semi-empirical erosion model which may be useful for scaling both terrestrial and lunar events.

**REFERENCES**

Alexander, J. D., W. M. Roberds, and R. F. Scott (1966). "Soil Erosion by Landing Rockets." Contract NAS9-4825 Final Report. Hayes International Corp., Birmingham, Alabama.

Carrier, W. David, III, Gary R. Olhoeft and Wendell Mendell (1991), "Physical Properties of the Lunar Surface," in *Lunar Sourcebook, A User's Guide to the Moon*, G. H. Heiken, D.T. Vaniman and B.M. French, eds. Cambridge University Press, Melbourne, Australia, pp. 475 – 594.

Immer, Christopher D., Paul Hintze, and Philip Metzger (2009). Manuscript in preparation.

Immer, Christopher D., John E. Lane, Philip T. Metzger, and Sandra Clements (2008). "Apollo Video Photogrammetry Estimation of Plume Impingement Effects," Earth and Space 2008, Long Beach, CA, Mar. 3-5, 2008.

Katzan, Cynthia M. and Jonathan L. Edwards (1991). "Lunar Dust Transport and Potential Interactions With Power System Components," NASA Contractor Report 4404. Sverdrup Technology, Inc., Brook Park, OH, pp. 8-22.

Lane, John E., Philip T. Metzger, and Christopher D. Immer (2008). "Lagrangian trajectory modeling of lunar dust particles," Earth and Space 2008, Long Beach, CA, Mar. 3-5, 2008.

Lucy, Paul (2009). Hawaii Institute of Geophysics and Planetology, University of Hawaii at Manoa. Personal communication.

Mason, Curtis C. (1970). "Comparison of Actual versus Predicted Lunar Surface Erosion Caused by Apollo 11 Descent Engine," *Geolog. Soc. of Am. Bull.* 81, 1807-12.

Mason, Curtis C. and E. F. Nordmeyer (1969). "An empirically derived erosion law and its application to lunar module landing," *Geolog. Soc. of Am. Bull.* 80, 1783-8.

Metzger, P.T., C.D. Immer, C.M. Donahue, B.T. Vu, R.C. Latta, III, M. Deyo-Svendsen (2009a). "Jet-induced cratering of a granular surface with application to lunar spaceports," *J. Aerospace Eng.* 22 (1), pp. 24-32.

Metzger, Philip T., Xiaoyi Li, Christopher D. Immer, and John E. Lane (2009b). "ISRU Implications of Lunar and Martian Plume Effects," AIAA 2009-1204, Proc. 47th AIAA Aerospace Sciences Conf., Orlando, FL, January 5-8, 2009.

Metzger, Philip T., Robert C. Latta III, Jason M. Schuler, and Christopher D. Immer (2009c). "Craters Formed in Granular Beds by Impinging Jets of Gas," Powders & Grains 2009 conf., Golden, Colorado, July 13-17, 2009.

Moore, H.J., R.E. Hutton, G.D.Clow, and C.R. Spitzer (1987). "Physical Properties of the Surface Materials at the Viking Landing Sites on Mars," U.S. Geological Survey Professional Paper 1389. Washington, USGPO, pp. 35-44.

Orpe, Ashish V., and D. V. Khakhar (2001), "Scaling relations for granular flow in quasi-two-dimensional rotating cylinders." *Phys. Rev. E* 64, 031302.

Park, Jaesung, Yang Liu, Kenneth D. Kihm, and Lawrence A. Taylor (2008). "Characterization of Lunar Dust for Toxicological Studies. I: Particle Size Distribution," *J. of Aerospace Eng.* 21(4), 266-271.

Roberts, Leonard (1963). "The Action of a Hypersonic Jet on a Dust Layer," IAS Paper No. 63-50, Institute of Aerospace Sciences 31st Annual Meeting, New York.